\documentclass[preprint,10pt,twocolumn,superscriptaddress,showpacs]{revtex4-1}
\usepackage{amsmath}
\usepackage{latexsym}
\usepackage{amssymb}
\usepackage{graphicx}
\usepackage[colorlinks=true, citecolor=blue, urlcolor=blue]{hyperref}
\usepackage{float}
\usepackage{amsfonts}

\begin{document}
\title{Certifiable randomness from temporal correlations}

\author{Shiladitya Mal}
\email{shiladitya@bose.res.in}
\affiliation{S. N. Bose National Centre for Basic Sciences, Salt Lake, Kolkata 700 098, India.}

\author{Manik Banik}
\email{manik11ju@gmail.com}
\affiliation{Physics and Applied Mathematics Unit, Indian Statistical Institute, 203 B. T. Road, Kolkata 700108, India.}

\author{Sujit K Choudhary}
\email{sujit@iopb.res.in}
\affiliation{Institute of Physics, Sachivalaya Marg,
Bhubaneswar-751005, Orissa, India}

\begin{abstract}
Leggett-Garg inequalities (LGI) are constrains on certain combinations of temporal correlations obtained 
by measuring one and the same system at two different instants of time. The usual derivations of 
LGI assume \emph {macroscopic realism per se } and \emph {noninvasive measurability}.
We derive these inequalities under a different set of assumptions, namely the assumptions
of \emph{predictability} and \emph{no signaling in time}. As a novel implication
of this derivation, we show that LGI can be used to certify randomness in 
a device independent way.

\end{abstract}


\maketitle

For testing the existence of superposition of macroscopically distinct
quantum states, Leggett and Garg \cite{legett} put forward the notion of {\it macrorealism}.
This notion rests on the classical paradigm \cite{benatti} that
\begin{itemize}
\item[(A1)] A macroscopic system with two or more macroscopically distinct 
states available to it will at all times be in one or the other of these states.
[\emph {Macroscopic realism per se }(MR)] 

\item[(A2)] It is possible, at least in principle, 
to determine which of these states the system is in, without any effect on the state itself or
on its subsequent dynamics. [\emph {Noninvasive measurability}(NIM)]
\end{itemize}
The conjunction of these two assumptions,  
namely the assumptions of  MR and 
NIM, together with the postulate of \emph {Induction} 
\cite{self, Leggett_1,Kofler,Kofler_1} give rise to Leggett-Garg inequalities (LGI).

Leggett-Garg inequalities have been a source of considerable research-interest 
\cite {ballentine,cliffton, foster91,  Nori}. 
However, there is still some controversy about the implication of its violation 
\cite{ballentine, legett87, legett88, maroney}.

In the present work, we give an alternative derivation of LGI by using different set of 
assumptions: the assumption of \emph{no signaling in time} (NSIT) and   
\emph{predictability}. The assumption of NSIT, described in \cite{Foster, Kofler_1}, says that
a measurement does not change the outcome statistics of a later measurement, whereas
\emph {predictability} is the assumption that one can predict the outcomes of all possible
measurements to be performed on a system \cite{wiseman}. This derivation, therefore,
allows us to conclude that in a situation where NSIT is satisfied, the violation of LGI
will imply the presence of ``true'' randomness. Apart from the theoretical significance,
this also has practical implications in the generation of Device-Independent (DI) 
certified randomness. 

From algorithmic information theory it is known that
randomness cannot be certified by any mathematical procedure \cite{Knuth}. 
The generation of randomness, therefore, must be
based on unpredictability of some physical phenomena, so that
the randomness is guaranteed by the inherent uncertain nature of the
physical theory. There is no such thing as true randomness 
in classical world as any classical phenomenon, can, in 
principle, be predicted. They appear random to us due to
lack of our knowledge and control of all the relevant degrees of freedom. 
Measurement on a quantum particle, on the other hand, is postulated
to give intrinsically random results.
The quantum measurements, therefore, can be used to generate 
true randomness \cite{qrng}.
But, for the reliability of the randomness thus generated, one needs 
to trust the devices which prepare and measure the quantum states.
Can randomness be certified in a Device-Independent way, i.e., can it be 
certified even without knowing the details of the devices used in its generation--
is a topic of current research interest \cite{Pironio, acin, scarani, banik}.
Interestingly, it can be certified in a DI way, provided, 
we consider the scenario which either involves different measurements on two
correlated particles \cite{wiseman, popescu94, Pironio, acin, scarani, banik} or which involve  
measurements on one and the same particle at different times. This latter is
the scenario for a Leggett-Garg test and is the subject matter of the present manuscript.

But, before moving to the Leggett-Garg test under the said assumptions of predictability
and NSIT, we, in the following, briefly describe the ontological framework of an operational
theory (for details of this framework, we refer to \cite{spek05,rudolph}), 
as this will subsequently be used in our derivation of LGI.   

 The goal  of an operational theory is merely to specify the probabilities $p(k|M,P,T)$ 
of different outcomes $k\in\mathcal{K}_M$ that may result from a measurement procedure 
$M\in\mathcal{M}$ given a particular preparation procedure $P\in\mathcal{P}$, and 
a particular transformation procedure $T\in\mathcal{T}$; where $\mathcal{M}$, $\mathcal{P}$ and 
$\mathcal{T}$ respectively denote the sets of measurement procedures, preparation procedures and 
transformation procedures; $\mathcal{K}_M$ denotes the set of measurement results for 
the measurement M. 

Whereas an operational theory does not tell 
anything about \emph{physical state} of the system, 
in an ontological model of an operational theory, the primitives 
of description are the actual state of affairs of the system. 
A preparation procedure is assumed to prepare a system with certain 
properties and a measurement procedure is assumed to reveal something 
about those properties. A complete specification of the properties of 
a system is referred to as the ontic state of that system. In an ontological 
model for quantum theory, a particular preparation method 
$P_{\psi}$ which prepares the quantum state $|\psi\rangle$, actually puts
the system into some ontic state $\lambda\in\Lambda$, $\Lambda$ denotes the ontic 
state space. An observer who knows the preparation $P_{\psi}$ may
nonetheless have incomplete knowledge of $\lambda$. Thus, in general, an 
ontological model associates a probability distribution 
$\mu(\lambda|P_{\psi})$ with preparation $P_{\psi}$ of 
$|\psi\rangle$. $\mu(\lambda| P_{\psi})$ is called the 
\emph{epistemic state} as it encodes observer's \emph{epistemic ignorance} about 
the state of the system. It must satisfy
\begin{equation}\nonumber
\int_{\Lambda}\mu(\lambda|P_{\psi})d\lambda=1~~
\forall~|\psi\rangle~\mbox{and}~P_{\psi}.
\end{equation}
Similarly, the model may be such that the ontic state $\lambda$
determines only the probability $\xi(k|\lambda,M)$, 
of different outcomes $k$ for the measurement method $M$.
However, in a deterministic model  $\xi(k|\lambda,M)\in \{0,1\}$.
The response functions $\xi(k|\lambda,M)\in[0,1]$, should satisfy
\begin{equation}\nonumber
\sum_{k\in\mathcal{K}_M}\xi(k|\lambda,M)=1~~\forall~~\lambda,~~M.
\end{equation}
 Thus, in the
ontological model, the probability $p(k|M,P)$ is specified as
\begin{equation}\nonumber
p(k|M,P)=\int_{\Lambda} \xi(k|M,\lambda)\mu(\lambda|P) d\lambda.
\end{equation} 
As the model is required to reproduce the observed frequencies (quantum predictions)
 hence the following must also be satisfied

\begin{equation}\nonumber
\int_{\Lambda} \xi(\phi|M,\lambda)\mu(\lambda|P_{\psi}) d\lambda = |\langle\phi|\psi\rangle|^2.
\end{equation} 

The transformation processes $T$ are represented by stochastic maps from ontic 
states to ontic states. $\mathcal{T}(\lambda'|\lambda)$ represents the probability
distribution over subsequent ontic states given that the earlier ontic state 
one started with was $\lambda$.

In a standard Leggett-Garg test, we consider a macroscopic object which 
is described by a set of macro variable $\{Q,Q', . . .\}$ whose values are considered
to be macroscopically distinct by some measure \cite{Leggett_1}.
In a series of runs, the object is prepared in the same initial state, and each 
preparation defines a new origin of time. 
Let us consider the case where macro variable 
$A\in\{Q,Q', . . .\}$ is measured at time $t_A (t_A > 0)$ and macro variable 
$B\in\{Q,Q', . . .\}$ at a later time $t_B$ \cite{self4}. The correlation function
$C_{t_At_B}\equiv \langle Q_{t_A} Q_{t_B}\rangle$ for measurements at $t_A$ and $t_B$
is obtained from the joint probability $P(A_{t_A}B_{t_B}|Q_{t_A}Q_{t_B})$ of obtaining
the results $A_{t_A}$ and $B_{t_B}$ from measurements of $Q$ at time $t_A$ and $t_B$ ($t_B> t_A$)
as
\begin{equation}\nonumber
 C_{t_At_B}=\sum_{A_{t_A}B_{t_B}} A_{t_A}B_{t_B} P(A_{t_A}B_{t_B}|Q_{t_A}Q_{t_B}).
\end{equation}
In the simplest case, the macro variable may obtain only two different values $\pm1$.
In such cases, \emph{macrorealism} together with \emph{induction} imply the LGI \cite{Leggett_1} of the 
Clauser-Horne-Shimony-Holt (CHSH) type \cite{chsh} $(t_1<t_2<t_3<t_4)$:
\begin{equation}\label{chshtype}
f^{LG}_4= -2\leq C_{t_1t_2}+C_{t_2t_3}+C_{t_3t_4}-C_{t_1t_4}\leq 2.  
\end{equation}
 or of the Wigner type \cite{wigner}
\begin{equation}\label{wignertype}
f^{LG}_3= -3\leq C_{t_1t_2}+C_{t_2t_3}-C_{t_1t_3}\leq 1
\end{equation}

In the ontological framework, the system's state is described by 
an ontic variable $\lambda$ and 
$P(A_{t_A},B_{t_B}|Q_{t_A},Q_{t_B}\lambda,\lambda \hspace{-.3cm}\rightarrow \hspace{-.3cm}\lambda')$ 
denotes the joint probability of obtaining outcome $A_{t_A}$ of measurement $Q_{t_A}$ 
performed at time $t_A$ 
and outcome $B_{t_B}$ of measurement $Q_{t_B}$ performed at 
a later time $t_B$; $\lambda \hspace{-.2cm}\rightarrow \hspace{-.2cm}\lambda'$ 
denotes the change of the system's ontic state conditioned 
that $A_{t_A}$ outcome has been obtained in 
measurement $Q_{t_A}$ at time $t_A$. The ontological model
then predicts for the observed probability as

\begin{eqnarray}
P(A_{t_A}B_{t_B}|Q_{t_A}Q_{t_B})&=&\int_{\lambda}\int_{\lambda'}d\lambda d\lambda' \mu(\lambda)\rho(\lambda'|Q_{t_A},A_{t_A},\lambda)\nonumber\\
&& P(A_{t_A}B_{t_B}|Q_{t_A}Q_{t_B},\lambda,\lambda \hspace{-.1cm}\rightarrow \hspace{-.1cm}\lambda'),
\end{eqnarray}

where $\mu(\lambda)$ and  $\rho(\lambda'|Q_{t_A},A_{t_A},\lambda)$ respectively
denote the distribution of the ontic variables 
prior to the measurement $Q_{t_A}$ and 
distribution of the ontic variables after obtaining the result $A_{t_A}$ 
in the measurement of $Q_{t_A}$. A crucial step in the derivation of LGI 
is to establish the following \emph{factorizability} relation which follows from 
the assumptions of \emph{macrolealism} and \emph{induction} \cite{Nori, Yearsley, self5}:
\begin{eqnarray}\label{factorisable}
P(A_{t_A}B_{t_B}|Q_{t_A}Q_{t_B},\lambda,\lambda \hspace{-.1cm}\rightarrow \hspace{-.1cm}\lambda')=\nonumber\\ P(A_{t_A}|Q_{t_A},\lambda)
P(B_{t_B}|Q_{t_B},\lambda)
\end{eqnarray}

It is noteworthy that, in contrast to \emph{macrorealism}, quantum mechanics predicts the outcome probability as:
\begin{equation}
P(A_{t_A}B_{t_B}|Q_{t_A}Q_{t_B})=\mbox{Tr}[\hat{\rho}(t_A)\hat{Q}_A]\mbox{Tr}[\hat{\rho}_{A_{t_A}}(t_B)\hat{Q}_B],
\end{equation}
where, $\hat{\rho}(t_A)$ is the quantum state of the system at time $t_A$, $\hat{Q}_A$ and $\hat{Q}_B$  
are the measurement operators for outcomes $A$ and $B$, and $\hat{\rho}_{A_{t_A}}(t_B)$ 
is the (reduced) quantum state at time $t_B$ given that at time $t_A$ result $A$ was obtained. 

For a two-level system undergoing coherent oscillations between the states 
with $Q =\pm1$, the optimal quantum violation of the inequality (\ref {chshtype}) 
is known to be $2\sqrt{2}$, whereas it is $\frac{3}{2}$ for the inequality (\ref {wignertype})
\cite{Barbieri}.

We, now, proceed to show that the assumptions of \emph{predictability} and 
\emph{no signaling in time} also lead to the factorizability condition
(\ref{factorisable}) and thus to the derivation of LGI.
 
NSIT \cite{Foster, Kofler_1, self2} is said to be satisfied if a measurement does 
not change the outcome statistics of a later measurement; i.e, 
$P(B_{t_B}|Q_{t_B})=P(B_{t_B}|Q_{t_A}Q_{t_B})$. 
Though \emph{macrorealism} implies both LGI as 
well as NSIT, the assumption of NSIT, alone, does not imply LGI.
However, it together with the assumption of \emph{predictability} imply LGI.
A model is said to be predictable if $P(A_{t_A}B_{t_B}|Q_{t_A}Q_{t_B})\in\{0,1\}$ 
for measurements at any time and for all measurement outcomes. 
As $P(A_{t_A}B_{t_B}|Q_{t_A}Q_{t_B})\in\{0,1\}$
hence conditioning on further variables cannot alter it, i.e.,
$P(A_{t_A}B_{t_B}|Q_{t_A}Q_{t_B},\lambda,\lambda \hspace{-.2cm}\rightarrow \hspace{-.2cm}\lambda')
=P(A_{t_A}B_{t_B}|Q_{t_A}Q_{t_B})$--no ontic variable further specify the probabilities. 
Now according to Baye's theorem, 
$P(A_{t_A}B_{t_B}|Q_{t_A}Q_{t_B})= 
P(A_{t_A}|B_{t_B}Q_{t_A}Q_{t_B})P(B_{t_B}|Q_{t_A}Q_{t_B})$. 
Again predictability implies $P(A_{t_A}|B_{t_B}Q_{t_A}Q_{t_B})=P(A_{t_A}|Q_{t_A}Q_{t_B})$. 
Assuming NSIT, we get  $P(B_{t_B}|Q_{t_A}Q_{t_B})=P(B_{t_B}|Q_{t_B})$ 
and due to \emph{Induction}, which says that 
measurement statistics at an earlier  
time should not depend on the what would be measured at a later time, we  
also have $P(A_{t_A}|Q_{t_A}Q_{t_B})= P(A_{t_A}|Q_{t_A})$. 
We, thus, have $P(A_{t_A}B_{t_B}|Q_{t_A}Q_{t_B},\lambda \hspace{-.1cm}\rightarrow \hspace{-.1cm}\lambda')
=P(A_{t_A}|Q_{t_A})P(B_{t_B}|Q_{t_B})$. The factorizability condition(\ref{factorisable}), then, follows
from conditioning the probabilities in the RHS on
$\lambda$  . 

The above derivation of LGI implies that either both or at least one of the underlying
assumptions is violated whenever LGI is violated. Imagine now a situation  
where LGI is violated but NSIT is satisfied. It would be worth mentioning here that 
NSIT is experimentally testable. 
In such situations, we can say that 
the model and hence the corresponding phenomena are not predictable.
Using the said situation, in the following, we 
show that temporal correlations are useful for DI randomness certification.    

\emph{Certifiable randomness from LGI:} 
Consider the standard Leggett-Garg scenario, where $\pm 1$-valued observables $Q_{t_A}$  and $Q_{t_B}$
are measured on a single system at two different times $t_A$ and $t_B$ respectively, where
 $t_A<t_B$. The joint probability distribution $P(11| Q_{t_A} Q_{t_B})$, of getting results  
$1$ at $t_A$ and $1$ at $t_B$ in such a scenario is calculated either by repeating the 
experiment many times or by employing an array of many identical systems. The other 
joint probabilities involved in Leggett-Garg inequalities (\ref{chshtype})
or (\ref{wignertype}) are calculated similarly to observe their violations. 
These probabilities are also analyzed to see whether NSIT is obeyed. In fact, 
in Ref. \cite{Kofler_1}, it has been shown that 
there exists probability distribution which violates LGI but satisfy NSIT. 
As we now know that such distribution cannot be predictable and therefore some randomness
is associated with it. The associated randomness can be quantified by min-entropy
\cite{Koenig} which is a statistical measure of the amount of randomness that 
a particular distribution contains. For a distribution $X$, it as defined as 
\begin{equation}\nonumber
H_{\infty}(X)\equiv \rm{log}_{2} \frac{1}{\max\limits_{x:\rm{Prob}(X=x)}\rm{Prob}(X=x)}
\end{equation}
\begin{figure}[t!]
\centering
\includegraphics[height=4cm,width=7cm]{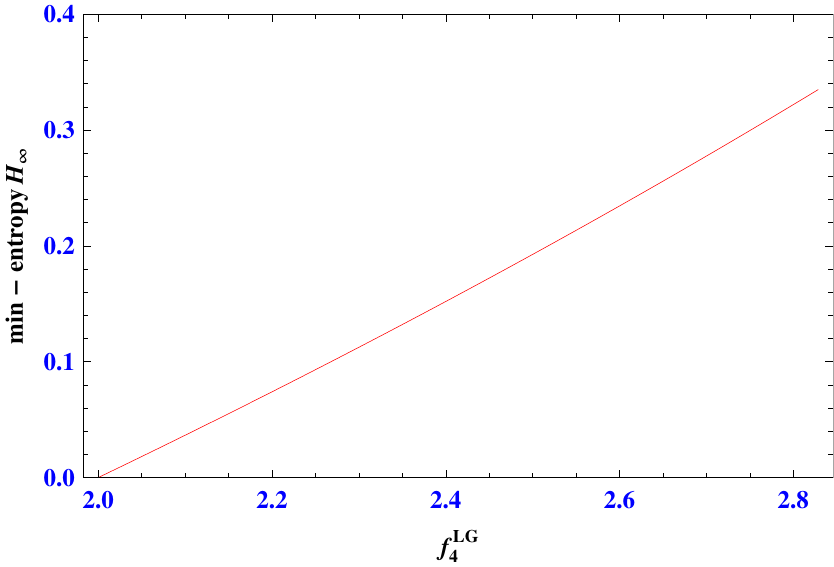}
\caption{(Color on-line) Certifiable randomness associated with Leggett-Gerg function $f^{LG}_4=f^{MR}_4+\epsilon$. Randomness is achieved for non zero value of $\epsilon$. }\label{fig1}
\end{figure}
Thus, to obtain the minimum amount of 
randomness associated with the violations of LGI (represented by
$f^{LG}_{\delta}=f^{MR}_{\delta}+\epsilon$, where $f^{MR}_{\delta}$ 
is the macrorealistic bound of $f^{LG}_{\delta}$; $\delta=3$ for inequality (\ref{wignertype}) and  
for (\ref{chshtype}) $\delta=4$; $\epsilon>0$),  
we need to first solve the following optimization problem:
\begin{eqnarray}\label{rand_lg}
P_{NSIT}(Q_{t_{\alpha}}, Q_{t_{\beta}})&=&\max\limits_{i,j}P(Q_{t_{\alpha}}=i, Q_{t_{\beta}}=j)\nonumber\\
&&\mbox{subject~to}~~f^{LG}_{\delta}=f^{MR}_{\delta}+\epsilon\nonumber\\
&& P(Q_{t_{\alpha}}=i, Q_{t_{\beta}}=j)\ge 0\nonumber\\
&&\sum_{i,j}P(Q_{t_{\alpha}}=i, Q_{t_{\beta}}=j)=1\nonumber\\
&& P(Q_{\mathcal{T}_{\alpha}},Q_{\mathcal{T}_{\beta}})~\mbox{satify~NSIT}. 
\end{eqnarray}
\begin{figure}[t!]
\centering
\includegraphics[height=4cm,width=7cm]{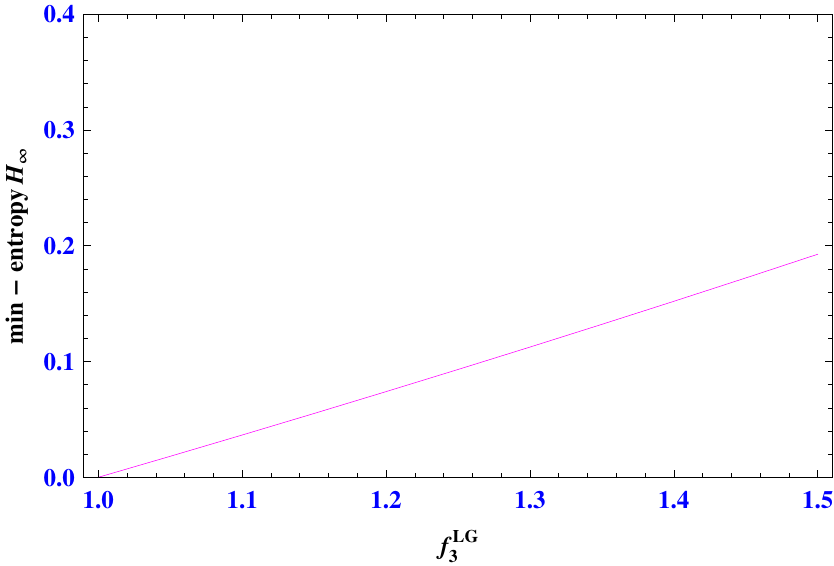}
\caption{(Color on-line) Certifiable randomness associated with Leggett-Gerg function $f^{LG}_3$.}\label{fig2}
\end{figure}
Having the optimized solution $P_{NIST}^*(Q_{t_{\alpha}}, Q_{t_{\beta}})$, 
the minimum randomness is calculated as 
$H_\infty(Q_{\mathcal{T}_{\alpha}},Q_{\mathcal{T}_{\beta}})=
-\log_2P_{NSIT}^*(Q_{t_{\alpha}}, Q_{t_{\beta}})$. 
We have considered the Leggett-Gerg function $f^{LG}_4$ and $f^{LG}_3$, one after another,
in the optimization problem (\ref{rand_lg}) and have numerically calculated
the minimum amounts of randomness with different values of $\epsilon$.  
We plot our findings in Fig.\ref{fig1} and Fig.\ref{fig2} \cite{Pironio_supply}.  

\emph{Concluding remarks}: Randomness is a valuable resource 
for various important tasks ranging from cryptographic applications 
to numerical simulations such as \emph{Monte Carlo} method (a useful 
technique which finds application in computational Physics, Statistical 
Physics, Physical Chemistry, Computational Biology, Computer Graphics, 
Finance and many other areas). For various such tasks, the genuineness of 
the used randomness is of primary concern. Thus, device independent certification 
and generation of randomness is very important from a practical point of view. 
Motivated by the work of Pironio and coworkers'\cite{Pironio}, many interesting results have 
been obtained, in recent times, in the field of DI certification and generation of randomness. 
All such methods use nonlocal correlations among spatially separated parties (which is guaranteed by 
Bell type inequalities violation \cite{scarani,Bell,Brunner}) to certify randomness. 
In this work, we have shown that temporal nonlocal correlations, i.e., correlations which violate 
Leggett-Gerg inequality, can also be used to certify randomness.
This work provides an important information 
theoretic application of LGI which can be implemented in laboratory with the present 
day's technology. From the perspective of experimental implementation the LGI-based DI 
randomness certification seems more feasible than its spatial analogue as  
it does not require entanglement \cite{self6}. Moreover, various successful experimental tests 
of LGI violation also give rise to the 
possibility towards further experiments with more macroscopicity involved. The present 
work also shows potential usefulness of such \emph{non-classical} 
macroscopic systems.  

This work is significant from an yet another perspective. Though Quantum Theory postulates
to have random measurement outcomes, it does not deny for a finer theory where the 
measurement outcomes are only apparantly random. In fact, there exists ontological models
which, in principle, can predict the outcomes of each individual mesaurement on a single particle
\cite{Bell,KS}. However, for really predicting the outcomes of a measurement, such a theory needs 
perfect knowledge of some variables. But, these variables are not accessible to the present day's
technologogy. Our analysis shows that even with some future technology (to which these variables 
are accessible and controllable), one cannot predict the outcomes of measurements performed on a 
single particle at two different times if the two-time correlations thus obtained 
violate LGI and satisfy NSIT.

\emph{Acknowledgment:} SKC acknowledges fruitful discussions with Pankaj Agrawal.
We also thank Guruprasad Kar for simulating discussions. 
SKC acknowledges support from the Council of Scientific
and Industrial Research, India, New Delhi.


\begin{thebibliography}{99}
\bibitem {self} In the present context, this postulate says that properties of an ensemble
are determined exclusively by initial conditions and cannot be affected by final conditions.
\bibitem {self4} One may of course choose $A=B$ and measure the same observable twice.
\bibitem {self5} Once the factorizability is achieved, the postulate of induction
is further used in calculating the correlation between measurement outcomes at 
two other different times. It allows one to freely choose the measurement times,
independent of the properties of the initially prepared state.
\bibitem{self2}The concept of NSIT has been describe in Ref.\cite{Foster} 
and for a comparative discussion between NIST and no-signaling 
(NS) see Ref.\cite{Kofler_1}, where it has been also argued that the assumption NSIT is a statistical version of the assumption NIM.
\bibitem {self6} However, the latter might have some other advantage over the LGI-based 
certification of randomness, especially for cryptographic purposes.
\end{thebibliography}
\end{document}